\documentclass[aps,prd,preprint,tightenlines,superscriptaddress,nofootinbib]{revtex4}
\usepackage{amsmath,amssymb}
\usepackage{graphicx,tabularx}

\def\bq{\begin{equation}}
\def\eq{\end{equation}}
\def\ba{\begin{eqnarray}}
\def\ea{\end{eqnarray}}
\newcommand{\sla}[1]{/\!\!\!\!#1}

\newcommand{\etal}{\textit{et al.}}

\newcommand{\mall}{m_{\tilde{t}_1,\tilde{t}_2,\tilde{b}_1,\tilde{b}_2}}

\begin{document}

\date{\today}

\title{Mixed top-bottom squark production at the LHC}

\author{D.~Berdine}
\email{berdine@pas.rochester.edu}
\affiliation{Dept. of Physics and Astronomy, University of Rochester, 
             Rochester, NY 14627}
\author{D.~Rainwater}
\email{rain@pas.rochester.edu}
\affiliation{Dept. of Physics and Astronomy, University of Rochester, 
             Rochester, NY 14627}

\begin{abstract}
We calculate cross sections for mixed stop-sbottom pair production at
the LHC, analogous to single-top production, a weak process involving
the $W$-$\tilde{t}_i$-$\tilde{b}_j$ vertex.  While coupling-suppressed
relative to QCD same-flavor squark pair production, the signal is
distinctive due to heavy-flavor tagging along with a possible
same-sign lepton pair in the final state.  SUSY backgrounds can often
be suppressed many orders of magnitude by taking advantage of distinct
kinematic differences from the signal.  Measuring the rate of this
process would add significant additional information to that gathered
from other SUSY processes.  If the stop and sbottom mixings can be
determined elsewhere, stop-sbottom production would provide for a
measurement of the weak squark gauge coupling and super-CKM vertex
factor.
\end{abstract}

\maketitle


\section{Introduction}
\label{sec:intro}

Spacetime Supersymmetry (SUSY) has become one of the leading candidate
extensions to the Standard Model (SM) of particle physics~\cite{SUSY}.
Besides the aesthetic appeal of maximally extending the Poincar\'e
group~\cite{Haag:1974qh}, the Minimal Supersymmetric Standard Model
(MSSM) also provides a natural weakly-interacting dark matter
candidate, a solution to the Higgs sector naturalness problem, and
likely unification of gauge couplings near the Planck scale, among
other features.

If SUSY exists, it must be a broken symmetry at low energy, as we do
not observe the opposite-spin-statistics partners of the SM field
content which would necessarily exist.  As a result, the squarks,
sleptons, charginos and neutralinos of the MSSM must be fairly massive
in comparison to their SM counterparts.  Previous high-energy physics
experiments such as LEP and Run~I of the Tevatron have put stringent
bounds on some of the sparticle masses.  It will fall to the CERN LHC
proton-proton collider to perform a conclusive SUSY search, although
the real physics would only begin were a potential SUSY discovery to
be made.

If LHC finds candidate SUSY particles, a great deal of work would be
required to prove or disprove the hypothesis that the new particle
content belongs to the SUSY partner spectrum of the SM content, and if
confirmed to perform a sufficient number of measurements to determine
the parameters of the SUSY
Lagrangian~\cite{Bechtle:2004pc,Lafaye:2004cn,Gjelsten:2005aw}.  For
the LHC the prospects of undertaking both tasks is uncertain, since
many of the weakly-interacting spartners might go unobserved, and many
of the colored spartners would have overlapping signatures;
disentangling these would be a daunting prospect.  The LHC will also
be hard pressed to measure sparticle masses: for some, only mass
differences may be
accessible~\cite{Hinchliffe:1996iu,Bachacou:1999zb,Allanach:2000kt,Gjelsten:2005aw}.
Performing such tasks as measuring the spins of new state, or
confirming that their couplings are indeed identical to the gauge and
Yukawa couplings of the SM, is largely unexplored territory
phenomenologically.  Ideally, a future linear collider (ILC) will be
constructed which could address many of these precision measurement
problems~\cite{ILC}, and there is considerable effort to understand
the synergy that would exist between LHC and
ILC~\cite{Weiglein:2004hn}.  However, there is also now the planned
luminosity upgrade to LHC, the SLHC~\cite{Gianotti:2002xx}.  While it
certainly won't be able to bring about a precision-era of physics and
guarantee disentanglement of whatever new physics is found at LHC, our
goal here is to explore just how far LHC and SLHC could push the
envelope in measurements of the new physics.

SUSY phenomenology at LHC has so far focused mainly on the dominant
$2\to 2$ processes, which for colored sparticles is QCD production of
gluino pairs, same-flavor squark pairs, and mixed gluino-squark pairs.
These production channels probe only the SUSY-QCD vertices.  While the
sparticle cascade decays will necessarily involve weak-interaction
vertices, the actual measurement is the number of events resulting
from the production cross section times branching ratio (BR) to a
given final state.  This alone is not enough information to separate
the different vertices.~\footnote{In principle, the QCD vertex could
be measured by observing all possible sparticle decays, so that the
total rate observed is $100\%$ of the BR.}  Analogous to the case of
single-top production in the SM~\cite{Chakraborty:2003iw}, observing
mixed-flavor squark production, which occurs via the weak vertex
$W$-$\tilde{q}_L$-$\tilde{q}_L'$ and semi-weak vertex
$W$-$g$-$\tilde{q}_L$-$\tilde{q}_L'$, would add additional information
about the sparticle interaction vertices.  Note that the weak
interaction, being left-handed, involves only the left-handed squark
partners of the quarks.  In fact, assuming that the various sparticle
BRs were determined in QCD pair production, mixed-flavor production
would then provide absolute measurements of the weak vertex couplings.

Weak production processes are na\"{\i}vely approximately two orders of
magnitude smaller than QCD production processes - although this also
depends on the relative mass spectrum with respect to which parton
luminosity dominates for that regime of $x$.  This would present a
problem for identifying this mixed-flavor signal.  There is some hope,
however.  In maximally-unifying scenarios, a.k.a.  ``GUT-inspired'',
over much of parameter space the colorless sparticles would be lighter
than the colored sparticles.  The cascade decay chains of squarks
would typically therefore proceed ultimately through the lightest
slepton (almost always a stau in popular unifying scenarios) or
lightest chargino, in either case giving a final-state
lepton~\footnote{A third possibility exists, where the cascade
proceeds ultimately through an on-shell $Z$ boson plus LSP.  This
scenario in general presents problems for SUSY analyses in
disentangling production processes.}.  Mixed-flavor production often
preferentially gives same-sign leptons in the final state, compared to
opposite-sign leptons from QCD pair production.  Since the
ATLAS~\cite{atlas_tdr} and CMS~\cite{cms_tdr} detectors at LHC will be
able to identify lepton charge sign to better than $10^{-4}$ accuracy,
this characteristic of weak mixed-flavor production could provide a
separation from the much greater rate of QCD production events.  Our
physics goal is to find and exploit characteristics of mixed-flavor
squark production which would make these weak cross sections stand out
against the much larger QCD squark production processes, providing a
way to measure the weak vertices.


\section{Stop-sbottom production at a hadron collider}
\label{sec:signal}

The best hope to look for mixed-flavor squark production is in the
third generation, i.e. top and bottom squarks.  The reasons for this
are threefold.  First, the largest first- and second-generation squark
pair production mechanism at LHC is typically
$uu\to\tilde{u}_L\tilde{u}_L$, which proceeds via a $t$-channel
Majorana gluino and dominates over squark-antisquark production
because of the very high initial-state valence $u$ quark luminosity.
This state would decay predominately to same-sign leptons via
$\tilde{u}_L\to\chi^+_1 d$, with the pair of lightest charginos,
$\chi^+_1$, decaying to same-sign leptons.  Second, the first- and
second-generation squarks are impossible to cleanly separate from each
other and from gluino pair and gluino-squark production, due to the
inability to tag light flavors.  Tagging the $b$ jets in stop-sbottom
mixed pairs would eliminate the first- and second-generation squark
contributions, up to the level of fake $b$ tag rates.  Third, because
gluinos are Majorana, their decay chains give equal probability for
either fermion sign~\cite{SSlep-go}.  As a result, gluino pairs and
squark-gluino production would be large sources of same-sign leptons.
Tagging $b$ jets can likewise reduce this fake source of same-sign
leptons, except that, depending on the spectrum, gluinos may decay
into stops or sbottoms themselves.  We will come back to these issues
later as background considerations.

For the third generation, the left- and right-handed squarks
$\tilde{q}_L$, $\tilde{q}_R$ can mix, forming mass eigenstates
$\tilde{q}_1$, $\tilde{q}_2$, with $\tilde{q}_1$ defined as lighter.
The mixing is proportional to the Yukawa couplings of the SM fermions.
The general mass matrix for stops is given by:
\ba
M_{\tilde{t}}^2 &=& 
	\left( 
	\begin{array}{cc} 
	M_{\tilde{t}_L}^2 & + m_t (A_t^* - \mu \cot\beta) \\
	  + m_t (A_t - \mu^* \cot\beta) & M_{\tilde{t}_R}^2 
	\end{array}
	\right)
\label{eq:sqmass}
\ea
where $\tan\beta$ is the ratio of Higgs sector vacuum expectation
values $v_u/v_d$, $\mu$ is the Higgsino mass parameter, $A_t$ is the
top squark trilinear scalar coupling in the soft SUSY breaking
potential, and
\ba
M_{\tilde{t}_L}^2 &=& m_{\tilde{t}_L}^2 + M^2_Z \cos 2\beta 
	\, (I_{3t} - s^2_W Q_t) + m^2_t \; ,
\label{eq:sq_diagonal_left}
\\
M_{\tilde{t}_R}^2 &=& m_{\tilde{t}_R}^2 + M^2_Z \cos 2\beta 
	\: s^2_W Q_t + m^2_t \; .
\label{eq:sq_diagonal_right}
\ea
where $m_{\tilde{q}_L}$, $m_{\tilde{q}_R}$ are the soft-breaking
masses in the SUSY Lagrangian.  For $b$ instead of $t$, replace
$\cot\beta$ with $\tan\beta$.  Because of this $L$-$R$ mixing in the
third generation, the SUSY weak vertices
$W$-$\tilde{t}_i$-$\tilde{b}_j$ are more complicated than the
corresponding $W$-$t$-$b$ vertex in the SM: they contain not just the
super-CKM angle $\widetilde{V}_{tb}$, but also the mixing angles of
the stops and sbottoms, which reflect the left-handed component of
each squark.  Expressed in terms of reduced parameters, the
$W$-$\tilde{t}_1$-$\tilde{b}_1$ coupling is
$g_W\widetilde{V}_{tb}\cos\theta_t\cos\theta_b$, where the mixing
angles are those that diagonalize Eq.~\ref{eq:sqmass} for stop and
sbottom squarks, respectively.  For $\tilde{q}_2$ instead of
$\tilde{q}_1$, $\cos\theta$ is replaced by $-\sin\theta$ to obtain the
left-handed component.  If SUSY were an exact symmetry,
$\widetilde{V}_{tb}\equiv V_{tb}$.  This is a very good approximation
even after SUSY breaking and evolution down to the electroweak
scale~\cite{Duncan:1983iq,Donoghue:1983mx}, although it can be altered
if SUSY breaking is not flavor-blind.

\begin{figure}[ht!]\begin{center}
\includegraphics[scale=0.8]{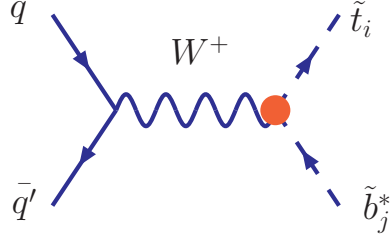}
\vspace{-2mm}
\caption{Feynman diagram for $s$-channel $\tilde{t}_1\tilde{b}^*_1$
production at a hadron collider.}
\label{fig:s-prod}
\vspace{-7mm}
\end{center}\end{figure}

\begin{figure}[ht!]\begin{center}
\includegraphics[scale=0.75]{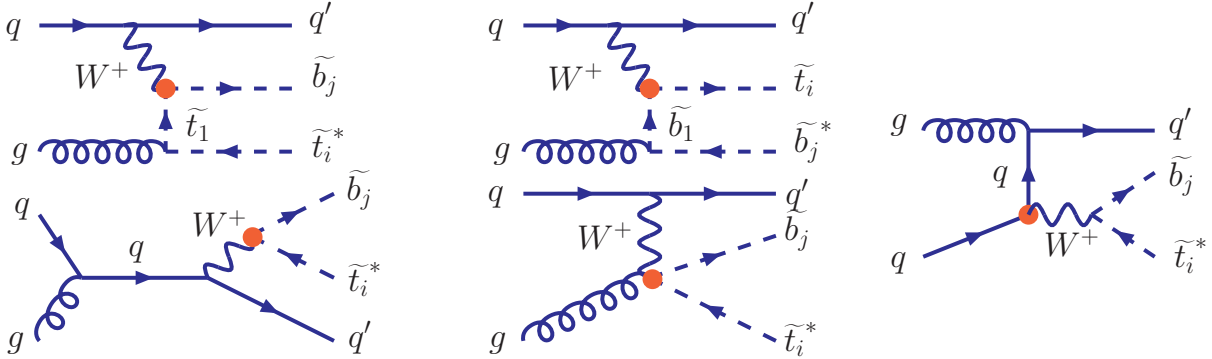}
\vspace{-8mm}
\caption{Feynman diagrams for $t$-channel $\tilde{t}_i\tilde{b}^*_j$
production at a hadron collider.  The dot highlights the weak vertex
to be measured.}
\label{fig:t-prod}
\vspace{-8mm}
\end{center}\end{figure}

As in SM single-top production, there are three ways to produce
mixed-flavor squarks at LHC: $s$-channel, $t$-channel, and
$W$-associated production.  The first two are shown in
Figs.~\ref{fig:s-prod} and~\ref{fig:t-prod}, respectively.  The third
is typically about 1/3 of the $t$-channel rate, but would be likely be
experimentally indistinct from QCD sbottom pair production.  We will
therefore ignore this channel.  The analogy to SM single-top
production is not complete, since in the SM case the light $b$ quark
mass allows it to be treated as an initial-state parton, which cannot
happen here.  The SM process also involves only one heavy state,
whereas here we have two, leading to different kinematics.  Finally,
in the SUSY case, because the squarks are scalars, there is a
four-point vertex which does not exist in the SM.  Although the graphs
of Fig.~\ref{fig:t-prod} are part of the real-emission QCD corrections
to the basic process of Fig.~\ref{fig:s-prod}, they behave very
differently.  As we will see, the $t$-channel cross sections are
always much larger than the $s$-channel.  This is a combination of the
larger suppression from an $s$-channel propagator at large squark pair
invariant mass, and initial-state gluon versus sea quark luminosity.

To calculate cross sections, we use matrix elements generated by a new
MSSM version of {\sc madgraph}~\cite{Stelzer:1994ta}, called {\sc
smadgraph}~\cite{SMG}.  Our calculations utilize CTEQ6L1 structure
functions~\cite{Pumplin:2002vw} for the incoming protons at LHC,
$\sqrt{s}=14$~TeV, and we choose the average of the two final-state
squark masses as the factorization and renormalization scales.  We
begin by analyzing the cross sections for the MSSM parameter space
benchmark point~\footnote{We use the SPS
benchmarks~\cite{Allanach:2002nj}, designed to represent a number of
canonical scenarios to aid exploratory phenomenology, only as examples
for convenience, not as any suggestion that these scenarios are more
likely to be realized in nature than any other.  They should be
regarded only as starting points for phenomenological investigation.} 
SPS1a~\cite{Allanach:2002nj}.  For this point, the squark masses are
$m_{\tilde{t}_1,\tilde{t}_2}=396,587$~GeV and
$m_{\tilde{b}_1,\tilde{b}_2}=517,547$~GeV.

As seen in Table~\ref{tab:SPS1a}, the LO total cross sections for
SPS1a are typically quite small, on the order of a few fb, although
even this would produce many hundreds of events for the planned
luminosity of 300~fb$^{-1}$ per experiment at the LHC, or thousands of
events for the 3000~fb$^{-1}$ per experiment for the planned
luminosity upgrade to the LHC, the SLHC~\cite{Gianotti:2002xx}
\begin{table}[htb]
\begin{tabular}{|c|c|c|c|c|}
\hline
& \; $\tilde{t}_{1}\tilde{b}_{1}$ \;
& \; $\tilde{t}_{1}\tilde{b}_{2}$ \;
& \; $\tilde{t}_{2}\tilde{b}_{1}$ \;
& \; $\tilde{t}_{2}\tilde{b}_{2}$ \\
\hline
\, $s$-ch., $\tilde{t}\,\tilde{b}^*$ \, & $1.3$ & $.18$ & $1.4$ & $.20$ \\
\hline
\, $s$-ch., $\tilde{t}^*\,\tilde{b}$ \, & $.51$ & $.07$ & $.51$ & $.07$ \\
\hline
\, $t$-ch., $\tilde{t}\,\tilde{b}^*$ \, & $4.7$ & $.73$ & $3.6$ & $.47$ \\
\hline
\, $t$-ch., $\tilde{t}^*\,\tilde{b}$ \, & $2.2$ & $.34$ & $1.6$ & $.21$ \\
\hline
\end{tabular}\centering
\caption{Cross sections [fb] for mixed stop-sbottom production at LHC
at MSSM benchmark point SPS1a.  For the $t$-channel results, we do not
impose a kinematic cut on the final-state quark.}
\label{tab:SPS1a}
\end{table}
\begin{table}[htb]
\begin{tabular}{|c||c|c|c|c||c|c|c|c|c|c|c|}
\hline
\, SPS \, & \;$m_{\tilde{t}_1}$ \; & \; $m_{\tilde{t}_2}$ \;
          & \;$m_{\tilde{b}_1}$ \; & \; $m_{\tilde{b}_2}$ \;
& \;\, $\tilde{t}\tilde{b}^*$ \;\, & \;\, $\tilde{t}^*\tilde{b}$ \;\,
& \; $\tilde{t}\tilde{t}^*$ \; & \; $\tilde{b}\tilde{b}^*$ \;
& \;\; $\tilde{g}\tilde{g}$ \;\;
& \, $\tilde{u}_L\tilde{g}$ \, & \, $\tilde{d}_L\tilde{g}$ \, \\
\hline
1a & 396 &  587 &  517 &  547 & 12.6   &  5.49  &  1630    & 560    & 4900 & 7370 & 3740 \\
1b & 653 &  839 &  778 &  828 &  1.25  &  0.468 &   95.9   &  44.7  &  238 &  582 &  260 \\
2  & 943 & 1318 & 1308 & 1567 &  0.045 &  0.014 &     7.74 &   0.74 &  780 &  128 &   52 \\
3  & 641 &  842 &  793 &  824 &  1.26  &  0.474 &   106    &  41.1  &  245 &  623 &  279 \\
4  & 540 &  696 &  619 &  691 &  3.74  &  1.52  &   306    & 168    & 1350 & 1890 &  897 \\
5  & 248 &  648 &  563 &  651 & 26.4   & 12.0   & 14200    & 280    & 1560 & 2780 & 1340 \\
6  & 476 &  704 &  640 &  662 &  3.90  &  1.59  &   575    & 164    & 1570 & 2510 & 1210 \\
7  & 802 &  894 &  861 &  878 &  0.620 &  0.223 &    31.8  &  25.1  &  218 &  527 &  234 \\
8  & 974 & 1091 & 1063 & 1082 &  0.167 &  0.055 &     8.16 &   5.68 &  500 &  400 &  174 \\
9  & 940 & 1152 & 1121 & 1278 &  0.139 &  0.046 &     8.94 &   2.76 &   43 &   85 &   34 \\
\hline
\end{tabular}\centering
\caption{Stop and sbottom masses [GeV], third-generation squark
LHC pair production LO cross sections [fb] for both mixed-flavor EW
and same-flavor QCD, and gluino pair and squark-gluino production, for
the various MSSM benchmark SPS points.  We combine $s$- and
$t$-channel cross sections for the signals, without any kinematic cut
on the final-state light quark in the $t$-ch. processes.  We also
combine all stop-sbottom combinations in the total EW rates, and the
$\tilde{q}_1\tilde{q}^*_1$+$\tilde{q}_2\tilde{q}^*_2$ rates for QCD.
For squark-gluino production, what we call $\tilde{u}_L$ is actually
$\tilde{u}_L$+$\tilde{c}_L$+$\tilde{d}^*_L$+$\tilde{s}^*_L$, which
together compose the contributions which would give a $\chi^+$ in
their decays; similarly for $\tilde{d}_L\tilde{g}$ production.}
\label{tab:SPS-all}
\vspace{-2mm}
\end{table}
(this would be, after all, a long-term measurement to determine the
SUSY Lagrangian parameters, not a discovery channel).  Note the
asymmetry in $\tilde{t}\tilde{b}^*$ v. $\tilde{t}^*\tilde{b}$
production.  It is due to the PDF asymmetry of the incoming protons,
LHC being a $p$-$p$ collider, and the dominance (by approximately a
factor of two) of initial-state up quarks over down quarks.
Table~\ref{tab:SPS1a} also reveals the left-handed and right-handed
components of the stops and sbottoms.  Since $\tilde{b}_1$ and
$\tilde{b}_2$ are nearly degenerate, the relative ratio of their cross
sections shows that it is $\tilde{b}_1$ which is mostly $\tilde{b}_L$.
In contrast, while $m_{\tilde{t}_2}\gg m_{\tilde{t}_1}$, the
$\tilde{t}_1$ and $\tilde{t}_2$ cross sections are nearly the same,
which shows that $\tilde{t}_2$ is more left-handed, although not by as
much of a margin as in the sbottoms.  In SUSY models where the mass
parameters are unified at the GUT scale, it is a general feature that
the lighter stop is more right-handed and the lighter sbottom is more
left-handed.  This comes about from the renormalization-group running
of the mass parameters from the GUT to the TeV scale due to the
different quantum numbers of top and bottom~\cite{Drees:1995hj}.

We give the total $\tilde{t}\tilde{b}$ cross sections in
Table~\ref{tab:SPS-all}, combining $s$- and $t$-channel results and
all $\tilde{q}_1$ and $\tilde{q}_2$ combinations, for all 10 SPS
points, along with the stop and sbottom masses at each point.  For
comparison, we also show the total
$\tilde{q}_1\tilde{q}^*_1$+$\tilde{q}_2\tilde{q}^*_2$ QCD production
rates, calculated here with {\sc smadgraph}, although this was first
performed in Ref.~\cite{Dawson:1983fw}.  The general feature is that
rates are lower for heavier squarks, as expected due to phase space
suppression, with an additional but non-obvious suppression from
mixings.  SPS1a has larger cross sections than most other SPS points,
except for the light-stop scenario of SPS5.  The questions are now,
given some rather small rates, might any of these produce enough
events to potentially be observable, what are the backgrounds in each
potentially viable scenario, and what would measuring these cross
sections tell us about SUSY?

Before addressing the signal and background rates, we make a few
comments on the significance of a potential stop-sbottom rate
measurement.  Ideally, one could measure the rate for each
stop-sbottom pair independently and with good accuracy.  One could
then extract each of $g_W\tilde{V}_{tb}\cos\theta_t\cos\theta_b$,
$g_W\tilde{V}_{tb}\cos\theta_t\sin\theta_b$,
$g_W\tilde{V}_{tb}\sin\theta_t\cos\theta_b$ and
$g_W\tilde{V}_{tb}\sin\theta_t\sin\theta_b$ independently.  The sum of
the squares of these is simply $g^2_W\tilde{V}^2_{tb}$, so this
combination would in principle be a test of SUSY, as it would measure
whether or not the weak $W$-$\tilde{t}$-$\tilde{b}$ vertex is of the
same strength as the SM weak vertex $W$-$t$-$b$.  Measurement of the
weak coupling strength in only a single channel, or couple of
channels, would be muddled by the mixing between left and right
states.  If this were all that was available, one would then argue
that the measurement aids determination of the mixing angles.

However, both of these ideas rely on the GUT-inspired model assumption
of flavor-diagonal SUSY soft-breaking masses.  This is not necessarily
strongly motivated, as presumably flavor physics enters at the GUT
scale if not before, and could well cause a deviation from such
unifying assumptions.  If the terms are not diagonal, then the mass
matrices which diagonalize all the squarks mix the first two
generations with the third, resulting in $\tilde{V}_{tb}\ne V_{tb}$,
and the stop-sbottom cross section(s) would help reveal the nature of
the SUSY soft-breaking terms.  Of course, in this case, all squarks
would have some decay branching fraction to heavy flavor, which may be
observable.  This is obviously a much more complicated scenario, which
we ignore in this first analysis for simplicity.


\section{Projections for some SUSY scenarios}
\label{sec:results}

Here we estimate the observability of stop-sbottom production in
various scenarios at LHC or its luminosity upgrade, the SLHC.  We
begin with the largest cross section of all the SPS points, SPS5, then
successively discuss SPS1a and various non-SPS points we formulate,
ranging from small variations on the SPS points to arbitrary
non-universal inputs scenarios motivated by phenomenology of the
low-scale MSSM spectrum~\cite{Drees:1995hj}.


\subsection{SPS5}

\subsubsection{Signal}

We examine the SPS5 scenario first, as it has the largest stop-sbottom
cross section of the SPS points.  This is due primarily to the lighter
stop being only 250~GeV, so the final state is not as phase space
restricted as in most of the SPS scenarios, which typically have much
more massive stops.  In SPS5, we see that $\tilde{t}_1$ always decays
as $\tilde{t}_1\to b\chi^+_1$, while the heavier stop decays this way
$16\%$ of the time, $57\%$ to $Z\tilde{t}_1\to Zb\chi^+_1$, and $20\%$
to $h\tilde{t}_1\to hb\chi^+_1$, while the remaining fraction goes to
$t\chi^0_{1,2}$.  The lighter sbottom decays mostly to
$W^-\tilde{t}_1$ ($77\%$), with $13\%$ to $t\chi^-_1$ and the
remaining fraction to $b\chi^0_2$.  The heavier sbottom has only a
rare $3\%$ decay to $t\chi^-_1$, preferring to go to $b\chi^0_1$ and
$W^-\tilde{t}_1$ approximately equally.  We summarize the relevant
sparticle masses, total widths at NLO in QCD, and branching fractions
relevant for this analysis in Table~\ref{tab:SPS5-BR}.

The decay combination we envision as being distinctive and indicative
of stop-antisbottom production is $\tilde{t}\to b\chi^+_1$ and
$\tilde{b}^*\to\bar{t}\chi^+_1$, yielding $b\bar{t}\chi^+_1\chi^+_1$
(and its charge-conjugate for antistop-sbottom production).  Having a
top quark to reconstruct (hadronically) in the final state is, we
feel, more distinctive than the $b\bar{b}\chi^+_1\chi^-_1W^\pm$ state
from sbottom decay to stop plus a $W$ boson, although this option
could be explored as it would enlarge the signal sample considerably
if accessible.  We ignore decays to neutralinos, because they are
charge-neutral and therefore cannot help distinguish $\tilde{q}$ from
$\tilde{q}^*$.

For SPS5, and typically in most minimal-supergravity (mSUGRA)
scenarios, the lightest chargino decays preferentially to a neutrino
plus the lightest slepton, typically the lightest stau,
$\tilde\tau^\pm_1$, as here.  The stau in turn decays to the lightest
SUSY particle (LSP), typically the lightest neutralino, plus a tau.
The detector signature would then be
$b\bar{b}jj\tau^\pm\tau^\pm+\sla{E}_T$.  Both $b$ jets would be tagged
by the vertex detector, with an efficiency of about $50\%$
each~\cite{atlas_tdr,cms_tdr}.  The taus can then in turn be
identified by their leptonic decay ($34\%$ BR with $\epsilon_{\rm
ID}=0.95$) or their 1-prong hadronic decay ($50\%$ BR and
$\epsilon_{\rm ID}=32\%$)~\cite{Rainwater:1998kj}.  For both cases,
the charge can be identified with uncertainty smaller than $10^{-4}$,
which is near-perfect, and sufficiently small to not worry about fake
same-sign events from opposite-sign backgrounds.  For the 1/4 of the
time in the SPS5 scenario that each chargino decays instead to LSP
plus $W$ boson, the leptonic decay of the latter would simply add to
the final state, although it would alter the expected ratio of $e$ or
$\mu$ to hadronic $\tau$ which would come from
$\tilde\tau\to\tau\chi^0_1$ decays only.  For LHC only, considering
600~fb$^{-1}$ by combining the results of two experiments, and
combining stau and tau decay modes, we could expect about 550 signal
events in the (++) channel before kinematic cuts, or 5500 at SLHC.
This is promising, but must be put in the context of visibility above
various backgrounds.

\begin{table}[ht!]
\begin{tabular}{|c|c|c|c|}
\hline
& \; $m$ \; & \, $\Gamma_{\rm NLO}$ \, & \, BR($q\chi^+_1$) \, \\
\hline
$\;\tilde{t}_1\;$   & 248 &  0.0365 & $100\%$  \\
$\;\tilde{t}_2\;$   & 648 & 14.4    & $17.3\%$ \\
$\;\tilde{b}^*_1\;$ & 563 & 14.4    & $14.3\%$ \\
$\;\tilde{b}^*_2\;$ & 651 &  0.635  & $ 3.9\%$ \\
$\;\tilde{u}_L\;$   & 680 &  6.56   & $66.0\%$ \\
$\;\tilde{d}^*_L\;$ & 684 &  6.46   & $65.4\%$ \\
\hline
\end{tabular}
\hspace{8mm}
\begin{tabular}{|c|c|c|c|c|}
\hline
& \; $m$ \; & \, $\Gamma_{\rm NLO}$ \, 
& \, BR($b\tilde{b}^*_1$) \, & \, BR($\bar{t}\tilde{t}_1$) \, \\
\hline
$\;\tilde{g}\;$ & 721 & 10.9 & $9.74\%$ & $26.6\%$ \\
\hline
\hline
& $m$ & \, $\Gamma_{\rm LO}$ \, 
& \, BR($\nu_\tau\tilde{\tau}^+$) \, & \, BR($W^+\chi^0_1$) \, \\
\hline
$\;\chi^+_1\;$ & 230 & 0.010 & $71.4\%$ & $28.6\%$ \\
\hline
\hline
& $m$ & \, $\Gamma_{\rm LO}$ \, 
& \, BR($\tau\chi^0_1$) \, & \\
\hline
$\;\tilde{\tau}^+_1\;$ & 184 & 0.297 & $100\%$ & \\
\hline
\end{tabular}
\caption{MSSM particle masses and widths [GeV], and important branching
ratios for scenario SPS5.  The LSP is the lightest neutralino,
$\chi^0_1$, with mass $120$~GeV.}
\label{tab:SPS5-BR}
\vspace{-2mm}
\end{table}

\subsubsection{Backgrounds}

There are multiple sources of backgrounds giving the same or similar
final states.  From the SM there is $t\bar{t}W^+(t\bar{t}W^-)$
production~\cite{Maltoni:2002jr}, which has a total cross section at
leading order (LO) of 290(136)~fb at LHC, or 2.47(1.16)~fb after decay
BRs to $b\bar{b}jj\tau^+\tau^++\sla{E}_T$, but before imposing any
kinematic cuts.  For a comparison with the case of signal taus
decaying hadronically these are the relevant cross sections, and
slightly larger than the signal, but within a factor of two.  If the
taus in the signal are to be observed in the lepton-hadron or dual
lepton modes, then the SM background would be a factor 4 larger,
although the signal would be a factor 5 larger.  After cuts the
$t\bar{t}W$ contribution is likely to be much smaller than the signal,
as $b$ jets and taus and the reconstructed top quark from stop-sbottom
decays will be produced much harder, i.e. with significantly more
transverse momentum in the detector.  More significantly, the SUSY
signal is likely to have significantly more missing transverse
momentum.  It is probable that one can impose cuts such that most of
the signal but very little of the SM background survives.  However,
this will require future complete decay simulation with kinematic
cuts.

QCD sbottom pair production can also give a similar final state:
$\tilde{b}_1\to W^-\tilde{t}_1\to W^-b\chi^+_1$ and
$\tilde{b}^*_1\to\bar{t}\chi^+_1$, yielding $b\bar{t}\chi^+_1\chi^+_1
+ W^-$.  The cross section for sbottoms pairs to this state of decays
is 20.7~fb at LO, compared to 3.0~fb for the signal.  This reveals a
weakness of our proposal, which is that if in a given scenario the
sbottom decays to stop with significant BR, then it could potentially
fake the signal.  However, this final state contains an extra $W$
boson, which could be vetoed in any of its decay modes.  Previous
applications of such a veto~\cite{Wveto} typically achieved more than
an order of magnitude reduction, but this applied to processes such as
top quark pair production, where the $W$ is produced with little
kinematic boost.  Here, the $W$ is the decay product of a 560~GeV
object decaying to the $W$ (80~GeV) plus a 250~GeV object: the $W$ is
likely to be boosted significantly, increasing the veto probability,
although an exact calculation with kinematic cuts on the final-state
$W$ decay products would be required to determine the precise
rejection efficiency.  Note that such a veto would also remove most of
the $W$-associated signal channel.  Hence, for this analysis we are
jusitifed in not considering it.  As for QCD stop pair production,
while it is large, it arises almost exclusively from
$\tilde{t}_1\tilde{t}^*_1$ pairs, which decay $100\%$ to opposite-sign
charginos.  We find that $\tilde{t}_2\tilde{t}^*_2$ decays to the
signal signature are much less than a $1\%$ background to the signal
rate.

Finally, SUSY QCD production of a first- or second-generation squark
plus a gluino, as well as gluino pairs, constitutes a potentially very
large background.  The reason is that squarks decay with large BR to
quark plus chargino, and gluinos in SPS5 are sufficiently massive to
decay to either top-stop or bottom-sbottom.  Because gluinos are
Majorana particles, they decay equally to particle-sparticle pairs of
either sign permutation.  This results in a very large cross section
for same-sign lepton pairs from squark+gluino and gluino pair
decays~\cite{SSlep-go}.  To give an example, the largest squark-gluino
cross section is $\tilde{u}_L\tilde{g}$, due to the initial state
containing a valence $u$ quark~\footnote{Note that $\tilde{q}_R$
production is not a background as right-handed squarks cannot decay to
charginos.}.  For SPS5 this alone has a cross section of 2300~fb at
LO, with NLO QCD corrections being a factor
1.3~\cite{Beenakker:1996ch}.  The relevant decays are $\tilde{u}_L\to
d\chi^+_1$ ($66\%$) and
$\tilde{g}\to\bar{t}\tilde{t}_1,b\tilde{b}^*_1$ ($27\%,10\%$).  These
decays produce identifiable final-state content identical to the
$t$-channel signal if the extra jet is observed: $j_h
b\bar{b}jj\tau^+\tau^++\sla{E}_T$, where $j_h$ represents a very hard
jet which arises from the $d$ quark.  This jet is highly energetic due
to the large mass difference between the squark and chargino, 680~GeV
v. 230~GeV.  Gluino pair production is 1600~fb at LO, with NLO QCD
corrections of a factor 1.9 for the SPS5 gluino and squark
masses~\cite{Beenakker:1996ch}.  The background arises from one gluino
decaying to quark plus squark (first- or second-generation only),
while the other decays as in the squark-gluino case.  The quark jet
from the first gluino's decay to quark plus squark, however, is
typically soft for SPS5, so ultimately the signature is essentially
the same as in squark-gluino production, but with a factor two for
gluino decay combinatorics, and another factor two for decay to either
up (down) or charm (strange) quark-squark pairs.  We would not propose
attempting to distinguish this background from the presence of the
extra soft jet, as an additional soft jet is likely to arise also in a
significant fraction of the signal simply due to QCD radiation.  Thus,
the heavy squark+gluino backgrounds are 3.8 and 1.9~pb for the (++)
and ($--$) cases, respectively.  These are a factor 240 larger than
the signal after squark and gluino decays to chargino plus sbottom.
Corrections to this from squark pair production with mistagged $b$
jets will be quite small, as the light quark rejection factor for $b$
jet fakes is 1/140.  We do not consider this latter source further
here.

\subsubsection{Differentiating signal and background}

\begin{figure}[ht!]\begin{center}
\includegraphics[scale=0.9]{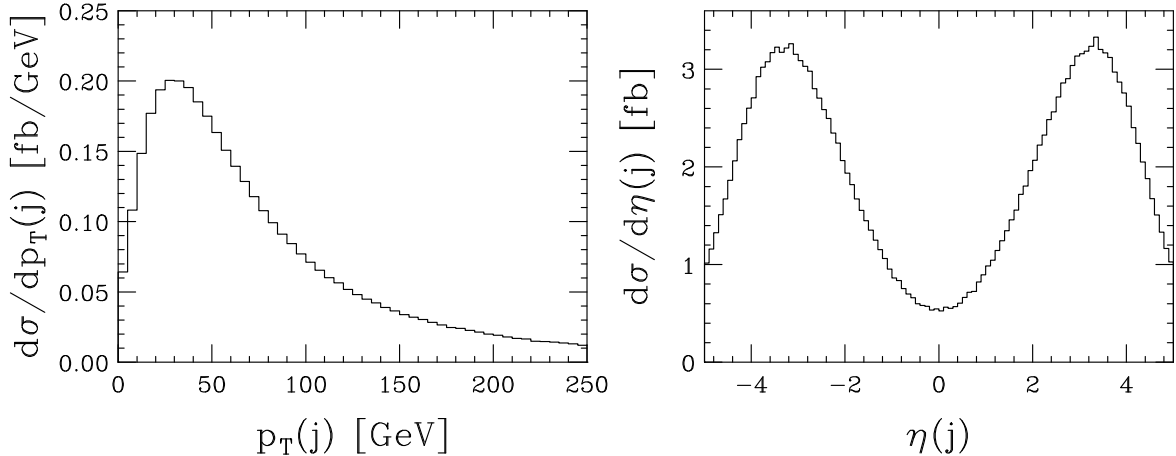}
\vspace{-4mm}
\caption{Transverse momentum and rapidity distributions of the extra
jet in $\tilde{t}_1\tilde{b}^*_1j$ production for the SPS5 scenario at
the LHC.  The ATLAS and CMS calorimeters cannot detect jets beyond a
rapidity of 5, or for $p_T\lesssim 20$~GeV.}
\label{fig:sigjet.SPS5}
\vspace{-7mm}
\end{center}\end{figure}

Fortunately, this heavy squark+gluino background can be suppressed
significantly by one of two means, either: (a) we take advantage of
the kinematic characteristics inherent in $\tilde{t}\tilde{b}^*j$,
where the extra jet tends to be emitted at large rapidity and high
transverse momentum in the detector, and require the hard jet from the
squark decay to be similarly far forward; or (b) we veto the extra
hard jet from the squark decay, since a sizeable fraction of the
signal does not have a high-$p_T$ jet.  We show the transverse momenta
and rapidity distributions of the extra jet in signal
$\tilde{t}_1\tilde{b}^*_1j$ production in Fig.~\ref{fig:sigjet.SPS5}
for illustration.
To be forward-tagged, we require $p_T(j)>30$~GeV and $3<|\eta(j)|<5$.
This provides a suppression factor of 50, while 1/3 of the signal
survives.  This would make the background approximately a factor 15
larger than the signal - a mediocre ratio, but not immediately to be
dismissed.  For a jet veto, which is more speculative and would
require far more in-depth investigation of additional QCD radiation in
the signal, we veto if $p_T(j)>40$~GeV and $|\eta(j)|<5$.  This veto
brings the background down by a factor 113, while $72\%$ of the signal
survives, for an overall signal-to-background (S/B) ratio of about
1/3.3, which is excellent.  The S/B ratio is superior in the jet veto
case because about twice as much signal survives, while the background
rejection due to the hard squark decay jet is more than twice as good.
Of course, our cuts values for these were not optimized, but serve as
a useful rough guide for the moment.

We also calculate the rates of $\tilde{b}\tilde{b}^*j$ production
using {\sc smadgraph}, and find that there is a factor 7 rejection of
this background if one requires a forward jet tag.  This would likely
be even better, as $\tilde{b}\tilde{b}^*jj$ is also quite large.
However, the rate for sbottom pair events to produce even one extra
jet anywhere in the fiducial volume of the detectors and with
$p_T(j)>40$~GeV is larger than either simple, parton-level
$\tilde{b}_i\tilde{b}^*_i$ rate at LO.  While there are large
uncertainties associated with such QCD calculations at LO, this is not
entirely unreasonable, because the NLO corrections to sbottom pair
production are quite large~\cite{Beenakker:1997ut}.  What our result
suggests is that the probability to produce an extra hard jet in
squark pair events is extremely large, which would only enhance the
rejection factor in the jet veto analysis.  A similar result holds for
squark+gluino production.  This issue is sufficiently complicated to
be beyond the scope of this paper, therefore is being investigated
elsewhere~\cite{XXnj}.

While we have not attempted to calculate the higher-order QCD (+1j)
contributions to the signal, we anticipate they are much smaller than
for the QCD processes, because no color is exchanged between the two
incoming partons.  Thus, additional radiation patterns should be
moderate in comparison to QCD.  This will be investigated in future
work~\cite{next}.  For now, our conclusion is that it is appropriate
to ignore the sbottom pair backgrounds, as their rejection factors due
to both hard QCD radiation and the extra $W$ boson certainly make them
very small corrections to the extremely large squark+gluino
background, as well as smaller than the signal.  We will therefore
focus on the squark+gluino backgrounds for our estimates of the
feasibility of detecting the signal.

\subsubsection{Numerical estimates}

It is beyond the scope of this first-state work to calculate the
signal and various backgrounds in full detail, with decays to a
detector final state (even at the parton level) and applying kinematic
cuts.  We relegate this next logical state to future work~\cite{next}
and here make only estimates of the upper bounds on signal cross
sections.  To make the estimates reasonable, we base them on branching
ratios to the desired final state, as well as the known detector
efficiencies for $b$ jet tagging and hadronic tau or lepton ID,
depending on the exact final state.

Our assumptions regarding the SLHC are that it collects 6000~fb$^{-1}$
(two experiments combined) and achieves the particle ID efficiencies
discussed previously.  We use our LO signal cross sections but NLO
squark+gluino background numbers, since the corrections are large;
this helps make our estimate conservative.  We assume that the
$t\bar{t}W$ background is eliminated completely, since it is the
smallest to begin with, and ignore the sbottom pair background as
discussed above.  Thus, the numbers of signal and background events in
the forward-tagged jet (jet veto) experiments are roughly 160(320) and
2350(1035).  The statistical significance for the forward jet tag case
is $\sim 3\sigma$, while for the jet veto option is $\sim 10\sigma$.
This corresponds to an approximately $12\%$ uncertainty on the cross
section from the veto analysis alone, half of that as the uncertainty
on the weak $W$-$\tilde{t}$-$\tilde{b}$ coupling.  Of course, this
assumes that $100\%$ of the signal would be retained after kinematic
cuts, which is obviously not correct.  We would nevertheless expect a
high retention rate because the final-state particles originate from
the decays of very heavy objects, thus will appear centrally and at
high transverse momentum in the detectors with a consequent high
probability to pass even conservative kinematic acceptance cut
requirements.  We would anticipate loss of signal due to cuts to be
less than a factor of two.  Thus, our estimate is cause for optimism
and invites more detailed investigation.  One should realize, too,
that the signal and background could behave differently with respect
to the final-state kinematic cuts, so there is some additional
systematic uncertainty at this stage as to the correct S/B ratio.


\subsection{SPS1a}

The benchmark point SPS1a has the next-largest stop-sbottom cross
section, about half the rate of the SPS5, as shown in
Table~\ref{tab:SPS-all}.  This is primarily because the lightest stop,
$\tilde{t}_1$, is more massive than in SPS5.  The QCD squark and
gluino pair background is much larger than in SPS5, in constrast,
because the other squarks and gluino are less massive than in SPS5.
However, production cross section numbers can be misleading because
the different spectrum results in different BRs of the various states.
In addition, the stop mixing is somewhat different, so that the
$\tilde{t}_2$ cross sections are comparable to $\tilde{t}_1$, as we
saw in Table~\ref{tab:SPS1a}.  Here, $\tilde{t}_1\to b\chi^+_1$ only
2/3 of the time, compared to always in SPS5, while
$\tilde{b}_1\to\bar{t}\chi^+_1$ occurs at three times the BR of SPS5.
In SPS1a, there are also non-trivial contributions to the same final
state from the sizeable $\tilde{t}_2$ rate.  In the backgrounds, while
there is far larger squark-gluino rate than in SPS5, the gluino BRs to
bottom-sbottom and top-stop are a factor of several smaller.  As with
SPS5, for SPS1a we summarize in Table~\ref{tab:SPS1a-BR} the sparticle
masses, total widths at NLO in QCD, and major branching fractions
relevant for this analysis.

We proceed as we did for the SPS5 scenario previously, constructing
two possible analyses, one demanding the presence of a far-foward
tagging jet, the other vetoing any additional hard jets, which the
heavy squark decays would give.  The rejection factors against the
squark-gluino backgrounds are 40 and 80, respectively, which are about
4/5 of the factors in SPS5.  Signal retention rates for both possible
analyses are similar to those in SPS5.

\begin{table}[htb!]
\begin{tabular}{|c|c|c|c|}
\hline
& \; $m$ \; & \, $\Gamma_{\rm NLO}$ \, & \, BR($q\chi^+_1$) \, \\
\hline
$\;\tilde{t}_1\;$   & 396 & 1.92  & $68.1\%$ \\
$\;\tilde{t}_2\;$   & 587 & 7.06  & $23.6\%$ \\
$\;\tilde{b}^*_1\;$ & 517 & 3.77  & $44.5\%$ \\
$\;\tilde{b}^*_2\;$ & 547 & 0.875 & $20.6\%$ \\
$\;\tilde{u}_L\;$   & 568 & 5.53  & $65.0\%$ \\
$\;\tilde{d}^*_L\;$ & 573 & 5.34  & $60.8\%$ \\
\hline
\end{tabular}
\hspace{8mm}
\begin{tabular}{|c|c|c|c|c|}
\hline
& \; $m$ \; & \, $\Gamma_{\rm NLO}$ \, 
& \, BR($b\tilde{b}^*_1$) \, & \, BR($\bar{t}\tilde{t}_1$) \, \\
\hline
$\;\tilde{g}\;$ & 607 & 4.33 & $10.4\%$ & $5.40\%$ \\
\hline
\hline
& $m$ & \, $\Gamma_{\rm LO}$ \, 
& \, BR($\nu_\tau\tilde{\tau}^+$) \, & \, BR($W^+\chi^0_1$) \, \\
\hline
$\;\chi^+_1\;$ & 181 & 0.0154 & $95.3\%$ & $4.7\%$ \\
\hline
\hline
& $m$ & \, $\Gamma_{\rm LO}$ \, 
& \, BR($\tau\chi^0_1$) \, & \\
\hline
$\;\tilde{\tau}^+_1\;$ & 136 & 0.155 & $100\%$ & \\
\hline
\end{tabular}
\caption{MSSM particle masses and widths [GeV], and important branching
ratios for scenario SPS1a.  The LSP is the lightest neutralino,
$\chi^0_1$, with mass $97.4$~GeV.}
\label{tab:SPS1a-BR}
\vspace{-2mm}
\end{table}

\begin{figure}[ht!]\begin{center}
\includegraphics[scale=0.8]{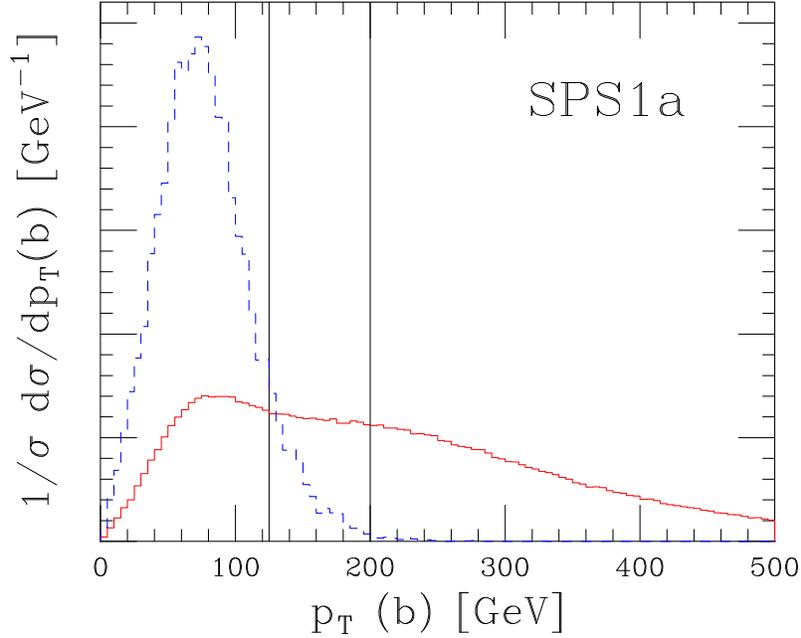}
\vspace{-2mm}
\caption{Transverse momentum ($p_T$) distribution of the $b$ jet which
does not come from a top quark decay, in the SPS1a scenario at the
LHC.  Signal $\tilde{t_1}\tilde{b}^*_1$ events are shown by the solid
red curve and $\tilde{u}_L\tilde{g}\to\chi^+_1 b\tilde{b}^*_1$
background events by the dashed blue curve.  The vertical lines are
two possible kinematic cuts: the left-hand cut would reduce the
background by a factor 10, with a $29\%$ loss of signal, while the
right-hand cut would yield a factor 270 background reduction, at a
cost of half the signal.}
\label{fig:pTb-SPS1a}
\vspace{-5mm}
\end{center}\end{figure}

At this point, we estimate after all BRs, detector ID efficiencies,
and including the NLO corrections for the heavy squark and gluino
backgrounds, that the forward jet tag analysis could gather up to
about 90 events in 6000~fb$^{-1}$, against a background of about 3200
events, while the jet veto analysis could retain up to about 220
signal events against a background of 1600.  Kinematic cuts on the
final state particles will reduce this somewhat, but again we argue
retention is likely to be highly efficient as the identifiable
particles come from the decays of very heavy objects and thus tend to
be highly boosted.  While the tagged analysis is quite poor, the veto
analysis already could achieve up to $\sim 5\sigma$, with up to a
statistical uncertainty on the signal cross section of about $20\%$.

\begin{figure}[ht!]\begin{center}
\includegraphics[scale=0.8]{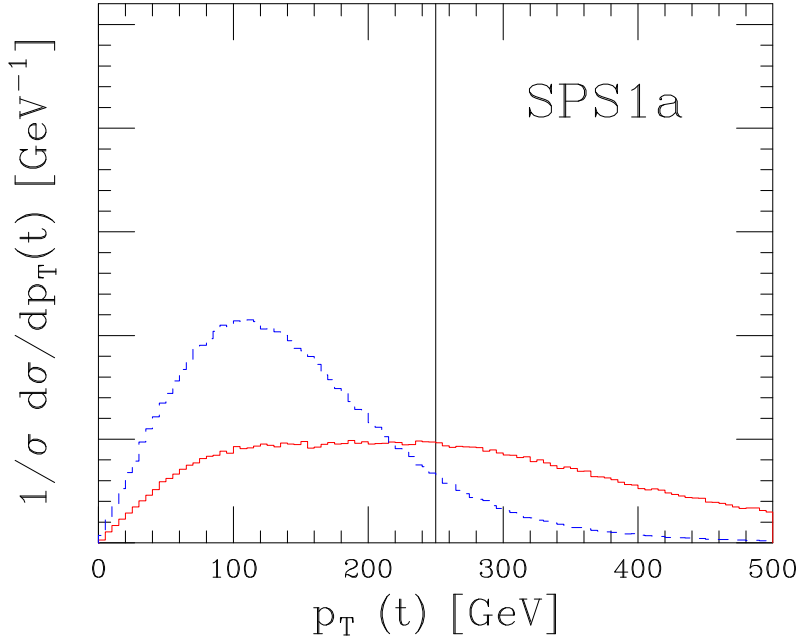}
\vspace{-2mm}
\caption{Transverse momentum ($p_T$) distribution of the top quark in
the SPS1a scenario at the LHC.  Signal $\tilde{t_1}\tilde{b}^*_1$
events are shown by the solid red curve and
$\tilde{u}_L\tilde{g}\to\chi^+_1\bar{t}\tilde{t}_1$ background events
by the dashed blue curve.  The vertical line represents a possible
kinematic cut which would eliminate $88\%$ of the background at a cost
of half the signal.}
\label{fig:pTt-SPS1a}
\vspace{-5mm}
\end{center}\end{figure}

One can improve on this by carefully considering the kinematic
characteristics of both signal and background.  The squark+gluino rate
can be broken down into two major parts: $\tilde{g}\to
b\tilde{b}^*$, and $\tilde{g}\to\bar{t}\tilde{t}$; they are roughly
equal.  The bottom quark jet in gluino decays, however, is extremely
soft, due to the small mass splitting between the gluino and the
sbottom.  The $b$ jet arising from stop decays in the signal, however,
is quite hard, due to the rather large mass splitting between stop
and chargino.

For this first-stage study we implement one level of decays with {\sc
smadgraph} and calculate the rates for the signal process
$\tilde{t}\tilde{b}^*\to b\chi^+_1\bar{t}\chi^+_1$ as well as the
background processes $\tilde{u}_L\tilde{g}\to d\chi^+_1
b\tilde{b}^*_1,d\chi^+_1\bar{t}\tilde{t}_1$.  We use the narrow width
approximation for phase space, as the sparticle widths are typically a
couple percent of their mass (see Table~\ref{tab:SPS1a-BR}), but
implement full matrix elements to retain spin angular correlations in
the backgrounds, using total widths for the various particles as
calculated at LO (to match the LO couplings in {\sc smadgraph}) by the
program {\sc sdecay}~\cite{Muhlleitner:2003vg}; we do consider overall
BRs at NLO, also given by {\sc sdecay}, since the corrections for the
gluino can be large~\cite{Beenakker:1996de}.  We show the $p_T$
distribution of the $b$ jet not arising from the top quark decay in
Fig.~\ref{fig:pTb-SPS1a}.  There is a dramatic difference in the
kinematic behavior, as expected~\footnote{We examined this also for
SPS5, but these kinematic features were not as dramatically different
there.}.  We display two possible cuts, one which reduces the
background by a factor 10 while retaining $71\%$ of the signal, the
other which totally obliterates the background (factor 270) at the
expense of fully half the signal.  Note that imposing such a cut would
also reduce the $\tilde{g}\to\bar{t}\tilde{t}$ rate by the
approximately the same amount as the signal, as the $b$ jet not from
top quark decay there comes from a stop as in the signal, so will
display a similar kinematic behavior.  A nice feature of the signal
distribution is its relative flatness to high $p_T$: if the tail of
the background distribution is wider than expected, one can adjust the
cut to compensate for this at very little relative cost to the signal.

Analogously for the other background contribution,
$\tilde{g}\to\bar{t}\tilde{t}$, the top quark daughter will tend to be
boosted more in the signal, where it was produced in conjuction with a
fairly light chargino, whereas in the background its decay partner is
a relatively heavy stop.  Although the differences are not as dramatic
as with the $b$ jet in stop v. gluino decays, we see in
Fig.~\ref{fig:pTt-SPS1a} that it is possible to find a cut which
reduces the background by about an order of magnitude while retaining
$50\%$ of the signal.  This is a somewhat higher price to pay than in
the gluino decay to bottom-sbottom, but it does improve the overall
statistical significance, as well as the more important
signal-to-background ratio.  Again, imposing this cut would reduce the
$\tilde{g}\to b\tilde{b}^*$ portion of the background similar to the
signal.

Taking both cuts together, we estimate up to about 32(78) signal
events could be retained in the forward-tag (jet veto) analysis, with
a background of about 210(105) events.  (Note that this is rather
pessimistic, since the two cuts are not truly orthogonal, thus the
actual signal retention rate should be higher.)  This improves the jet
veto case to a very respectable S/B ratio of 1/1.3, up to about
$7.6\sigma$ statistical significance, and potentially up to a $20\%$
statistical uncertainty on the signal cross section.  This is again
idealized, but we would expect a high signal retention rate after
cuts, as most of the final-state particles will be highly boosted and
register well in the detectors.  Again, as with SPS5 there is some
systematic uncertainty on S/B that arises from not knowing whether the
signal and background behave similarly for the final-state kinematic
cuts.  Further study will clarify this.


\subsection{General MSSM scenarios}

While the SPS scenarios constitue a useful framework in which to study
SUSY phenomenology at upcoming collider experiments, we remind
ourselves that they do not by any means fully represent the plethora
of possible MSSM parameterizations.  They are only starting points:
the SUSY breaking schemes mSUGRA, gauge-mediated (GMSB) and
anomaly-mediated (AMSB) all make broad assumptions about unification
of input mass parameters at some high scale.  While the motivations
for doing so are elegant, they are not by any stretch of the
imagination certain.

Based on our experiences studying SPS5 and SPS1a, we can draw a few
simple conclusions to guide exploration of parameter space with a more
open mind.  First, and most obviously, LHC has the potential to
observe stop-sbottom production when the cross sections are large
enough to produce a statistically useful number of events; the
backgrounds in the SPS cases we examined so far can be suppressed to
approximately the level of the signal.  For the production of heavy
objects, LHC is limited mostly by phase space.  Thus, scenarios with
lighter stops and sbottoms are more desireable.

Second, very large backgrounds arise from first- and second-generation
squark+gluino and gluino pair production, but they can also be heavily
suppressed using additional kinematic information in the events.
Scenarios with very heavy squarks and gluinos relative to stops and
sbottoms are in some sense variants of SPS5 and highly likely to be
accessible by comparison.  Alternatively, if the gluino is lighter and
cannot decay to stops or sbottoms, then this background completely
disappears.  The other squarks cannot become a sizeable background on
their own, as their decays would not yield heavy flavor quarks
(including a top quark) in the final state.

One does have to worry about gluinos lighter than the sbottom,
however, as the strong decay $\tilde{b}\to b\tilde{g}$ then becomes
possible, which tends to dominate in BR over the weak decays to bottom
quark plus chargino.  Thus, while there is no background, there is
little signal rate left to the same-sign charged lepton pair we
advocate.  Were $b$ jet-charge tagging to be possible with decent
efficiency, then such a scenario might be accessible.  One could also
consider the possibility of identifying mixed-flavor production in
this scenario by observing a single lepton from the stop decay, two
$b$ jet tags, and a characteristic gluino, thus separating the signal
from stop pair or sbottom pair production; but this is rather
speculative and would require a completely different analysis.

In searching for other viable scenarios, we do not rigorously check
for known constraints from present data, except for the Higgs and
chargino limits from
LEP~\cite{Barate:2003sz,LEP-MSSM-H,Clerbaux:2003gq,Degrassi:2002fi},
and of course the requirement of a neutral, colorless LSP.  Nor do we
attempt a thorough, systematic exploration of parameter space at this
early stage.  Our goal is simply to emphasize how different MSSM
realizations could change expectations for the stop-sbottom signal
qualitatively.  An early warning: scenarios with a small NLSP-LSP mass
difference can give final-state taus or leptons which are soft a large
fraction of the time, presenting a rate problem after detector
acceptance cuts.  The exception to this is when the mass difference is
so small that the NLSP is long-lived, as we will see below.

\begin{itemize}

\item[-] \underline{Lighter stop and sbottom via increased mixing.}
Starting with SPS1a and simply increasing the value of $A_0$ to
-500~GeV, thus maintaining the unification of input mass parameters,
results in about a $70\%$ cross section increase for the slightly
lighter stops and sbottoms.  However, the gluino mass doesn't change,
while its BR to top-stop increases by a factor three.  Signal 
significane is likely to decrease slightly.

\item[-] \underline{Heavier gluino via increased $M_3$.}
Starting with the low-energy effective SPS1a point but instead
allowing non-universal values for $m_{1/2}$, setting
$M_{1,2,3}=100,200,500$~GeV and $A_{t,b}=-2$~TeV, we obtain a very
heavy gluino, $m_{\tilde{g}}=1154$~GeV, while the stop and sbottom
masses similar to those of SPS1a, and their total production cross
sections about 1/3 the SPS1a rate.  Now the squark+gluino backgrounds
are only a few hundred fb to start, almost two orders of magnitude
smaller, but with BRs to stops and sbottoms a factor of a few larger
than SPS1a.  With this spectrum the kinematic tricks we identified for
SPS1a will not work as well, as the gluino-sbottom mass splitting is
now several hundred GeV.  However, the overall lowering of background
from production rate, less the increased BRs, is about the same as the
suppressed rate of SPS1a via kinematic differences.  Such a scenario
looks promising only if the gluino BR to stop or sbottom and loss of
kinematic cut effectiveness does not grow at a faster rate than the
cross section falls off due to phase space.

\item[-] A variant of the above with $M_{1,2,3}=150,300,1000$~GeV and
universal $m_0=500$~GeV at the TeV scale predicts a 970~GeV gluino, a
light stop of 322~GeV and lightest sbottom of 475~GeV.  The signal
rates are similar to SPS5, with a smaller background than in SPS5,
from both smaller production rates and smaller gluino decay rates to
stops and sbottoms.  We find a wide parameter space around this
scenario which produces a similar spectrum.


\item[-] \underline{Gluino lighter than the sbottoms.}
Starting at SPS1a but allowing non-universal masses, we lower
$m_{1/2}$ for $SU(3)$ only to 100~GeV, set $A_0=0$ and
$m_{\tilde{t}_R}=m_{\tilde{b}_R}=150$~GeV at the GUT scale.  The
gluino mass is now 271~GeV, while the lighter sbottom mass is 274~GeV.
The lightest stop is 193~GeV; the heavier stop and sbottom are both
under 400~GeV.  The total of stop-sbottom cross sections is 160~fb,
with no background from squark+gluino pairs.  However, the lighter
sbottom does not decay at all to chargino, rather mostly to $b$+LSP.
The $\tilde{b}_2$ actually dominates the production rate, and it has a
$6\%$ BR to chargino.  The rate to $b\bar{t}\chi^+_1\chi^+_1$ is
larger than at SPS1a by a small factor.  The chargino is lighter than
the stau in this scenario, which would decay like a $W$ boson, with an
effective overall ID efficiency via its decays to $e,\mu$ about that
same as a stau pair.  Such scenarios look extremely promising.  Other
variations on this theme, with $m_{\tilde{g}}\sim m_{\tilde{b}}$, are
easy to find.


\end{itemize}

Our qualitative survey suggests that parameterizations which give
larger signal cross sections are often good, but also sometimes less
viable, for a variety of reasons.  Usually the culprit is either a
dramatically larger BR of the gluino to a stop or sbottom, or the stop
and sbottom BRs themselves squeeze the distinctive signal we propose.
Analogously, smaller cross sections can sometimes be deceptively good,
often for similar reasons.

\bigskip
\underline{\it Stau coannihilation scenario}
\smallskip

We examine a stau-coannihilation
scenario~\cite{Ellis:1998kh,Ellis:1999mm}, inspired by the dark matter
relic density~\cite{Bertone:2004pz}.  We choose TeV-scale input
parameters of $M_{1,2,3}=100,100,800$~GeV, $m_0=800$~GeV for the first
two generations, $m_{\tilde\tau}=127$~GeV, $m_{Q_L}=m_{t_R}=400$~GeV,
$m_{b_R}=300$~GeV and $A_{t,b}=-700$~GeV.  The stau mass obtained is
100~MeV above the LSP at 97~GeV, with a 101~GeV chargino - this is the
principal features of this scenario.  As a consequence, the stau is
relatively long-lived.  We selected other parameters to guarantee
reasonably light third-generation squarks: the lighter stop and
sbottom are 243 and 288~GeV, respectively, with their heavier partners
at 544 and 406~GeV.  The other squarks and gluino are somewhat heavier
at 810~GeV.  With regard to the colored sparticle sector, this point
is not all that different from SPS5.

The largest signal cross section is $\tilde{t}_1\tilde{b}^*_2$ at
67~fb, as shown in Table~\ref{tab:stauco-xsec}.  The stop decays
$100\%$ to our desired chargino, and the sbottom 1/4 of the time.  The
chargino in turn decays exclusively to the lighter stau, which is
long-lived.  Herein lies the munificence: the detection efficiency for
each stau will be extremely close to $100\%$, with perfect charge
identification.  Already, then, any such scenario will achieve at
least a factor 4 detection efficiency improvement over any model where
the stau decays promptly to a tau.  Moreover, there is no SM
background at all.  For a jet tag strategy, we calculate 2.63~fb for
the signal final state $b\bar{t}\tilde\tau^+_1\tilde\tau^+_1$ where we
include a factor 2/3 for the hadronic BR of the top quark.  For a jet
veto strategy, we calculate 4.13~fb.

Squark+gluino backgrounds are similar to SPS5, but with $50\%$ BR of
gluino to bottom-sbottom and top-stop of the (++) charge sign, larger
than SPS5.  As implied, the gluino does not decay to any first- or
second-generation squark plus quark, thus removing gluino pair
production as a background.  Rejection factors for the jet tag and jet
veto scenarios are extremely good: 60 and 200, respectively.  By the
time NLO corrections, jet tag/veto rejection factors and BRs are taken
into account, we find about 2.09~fb for the jet tag case and 0.61~fb
using instead a jet veto.  These cross sections are smaller than the
signal.

However, the real background comes from QCD sbottom pair production.
Specifically, $\tilde{b}_2\tilde{b}^*_2$ with a NLO cross section of
approximately 1.9~pb, as $\tilde{b}_2\tilde{b}^*_2$ cannot fake the
signal at this point.  After BRs, this is about 170~fb, although this
is comparable to the sbottom pair rate for SPS5.  The difference here
is that the squark+gluino background is smaller, rather than
significantly larger.  Again we assume that there is an order of
magnitude rejection of sbottom pairs via the extra $W$ boson, and
another factor of 5 from a jet veto on the extra radiation.  There is
a large uncertainty associated with this, but as we will see, it is
largely irrelevant for observational success at this point.

\begin{table}[htb]
\begin{tabular}{|c|c|c|c|c|}
\hline
& \; $\tilde{t}_{1}\tilde{b}_{1}$ \;
& \; $\tilde{t}_{1}\tilde{b}_{2}$ \;
& \; $\tilde{t}_{2}\tilde{b}_{1}$ \;
& \; $\tilde{t}_{2}\tilde{b}_{2}$ \\
\hline
\, $s$-ch., $\tilde{t}\,\tilde{b}^*$ \, & $0.69$ & $9.80$ & $0.11$ & $2.19$ \\
\hline
\, $s$-ch., $\tilde{t}^*\,\tilde{b}$ \, & $0.32$ & $4.33$ & $0.05$ & $0.84$ \\
\hline
\, $t$-ch., $\tilde{t}\,\tilde{b}^*$ \, & $2.28$ & $57.2$ & $0.95$ & $8.18$ \\
\hline
\, $t$-ch., $\tilde{t}^*\,\tilde{b}$ \, & $1.21$ & $28.6$ & $0.44$ & $3.74$ \\
\hline
\end{tabular}\centering
\caption{Cross sections [fb] for mixed stop-sbottom production at LHC
in a stau-coannihilation scenario.  For the $t$-channel results, we do
not impose a kinematic cut on the final-state quark.}
\label{tab:stauco-xsec}
\end{table}

\begin{table}[htb!]
\begin{tabular}{|c|c|c|c|}
\hline
& \; $m$ \; & \, $\Gamma_{\rm NLO}$ \, & \, BR($q\chi^+_1$) \, \\
\hline
$\;\tilde{t}_1\;$   & 243 & 0.87 & $100\%$  \\
$\;\tilde{t}_2\;$   & 544 & 10.7 & $26.0\%$ \\
$\;\tilde{b}^*_1\;$ & 288 & 0.19 & $6.02\%$ \\
$\;\tilde{b}^*_2\;$ & 406 & 6.30 & $25.8\%$ \\
$\;\tilde{u}_L\;$   & 806 & 9.3  & $66.0\%$ \\
$\;\tilde{d}^*_L\;$ & 810 & 9.2  & $64.1\%$ \\
\hline
\end{tabular}
\hspace{8mm}
\begin{tabular}{|c|c|c|c|c|}
\hline
& \; $m$ \; & \, $\Gamma_{\rm NLO}$ \, 
& \, BR($b\tilde{b}^*$) \, & \, BR($\bar{t}\tilde{t}$) \, \\
\hline
$\;\tilde{g}\;$ & 810 & 31.4 & $30\%$ & $20\%$ \\
\hline
\hline
& $m$ & \, $\Gamma_{\rm LO}$ \, 
& \, BR($\nu_\tau\tilde{\tau}^+$) \, & \\
\hline
$\;\chi^+_1\;$ & 101 & 0.00115 & $100.0\%$ & \\
\hline
\end{tabular}
\caption{MSSM particle masses and widths [GeV], and important branching
ratios for a dark matter-motivated stau coannihilation scenario with
squark and gluino masses similar to SPS5, but with a slightly lighter
$\tilde{b}_2$.  The LSP is the lightest neutralino, $\chi^0_1$, with
mass $97.0$~GeV.  The stau is the NLSP, with mass $97.1$~GeV, and is
long-lived on detector timescales..}
\label{tab:stauco-BR}
\vspace{-2mm}
\end{table}

Because of the anticipated large detection efficiency for long-lived
staus, about a factor 5 over staus which would promptly decay to taus,
our rough estimate of observability here will be for the LHC already,
rather than the SLHC.  Using a jet veto strategy we would anticipate
approximately 620 double-$b$-tagged signal events (including hadronic
top decay), on top of a total background of about 400 events; S/B
would be better than 1/1.  This would provide for a remarkable
$>30\sigma$ detection and statistical uncertainty on the signal cross
section to this final state of about $5\%$.  Even if the background
rejection factors due to $W$ or jet veto turn out to be off by an
order of magnitude, this would still allow for a $10\sigma$
observation and up to an $11\%$ statistical uncertainty on the cross
section.

Needless to say, at SLHC this channel would rapidly become dominated
by systematic errors, such as the QCD uncertainty on the signal
theoretical cross section or the measurement of the
$\tilde{b}_2\tilde{b}^*_2$ rate and its BRs.  However, because the
dominant background is likely this squark pair rather than
squark+gluino production, there is an additional handle one could
apply to improve the situation: the signal has an approximately 2:1
asymmetry for (++) v. ($--$) production, due to the predominance of
initial-state valence $u$ over valence $d$ partons.  QCD sbottom pair
production is totally symmetric, in contrast.  Thus, by measuring the
charge asymmetry of the final state,
\bq\label{eq:asym}
A \; = \; \frac{\sigma_{++}-\sigma_{--}}{\sigma_{++}+\sigma_{--}} \; ,
\eq
one could gain additional strong leverage over residual systematic
uncertainty of the QCD background.

\bigskip
\underline{\it GMSB scenario}
\smallskip

A particularly interesting scenario is gauge-mediated SUSY breaking
(GMSB), which has a gravitino LSP.  Because the decay of the NLSP
would be of gravitational strength, the NLSP is typically long-lived,
and if charged would be extraordinarily noticeable in experiment as a
heavy charged object passing through the muon chambers.  Because of
this, the efficiency to detect SUSY events would be very close to
$100\%$ if the NLSP is charged, completely eliminating SM backgrounds
and retaining nearly the whole rate of stop-sbottom pairs, modulo the
efficiency for $b$-tagging, needed to identify that the SUSY signal
comes from third-generation squarks.

SPS7 is a GMSB scenario with stau NLSP, but because of the very large
stop and sbottom masses, there is practically no rate for mixed-flavor
pairs at LHC.  We instead look for parameterizations of GMSB where the
stops and sbottoms are light enough to be produced at the LHC, and
also where the gluino is slightly lighter than the sbottoms, so that
the BR($\tilde{g}\to b\tilde{b}$) does not dominate.  SUSY backgrounds
are then composed only of light squarks and gluinos which decay to
same-sign staus, with light jets mistagged as $b$ jets.  The large
suppression of such fakes may make such a scenario feasible for much
smaller signal cross sections than in SPS1a or SPS5.

We easily find an example of such a parameterization, by choosing
$M_{mes}=100$~TeV, $M_{SUSY}=20$~TeV, $\tan\beta=5$, $\mu<0$, and the
presence of 3 lepton and 2 quark messenger states.  The latter is a
somewhat odd choice, but is not restricted in any way.  For these
choices, $\mall=350,410,360,370$~GeV.  Only the $\tilde{b}_2$ cross
sections are of decent size: that for $\tilde{t}_1\tilde{b}^*_2$ is
about 1~fb and that for $\tilde{t}_2\tilde{b}^*_2$ is about 3.6~fb,
both obtained by requiring the forward-tagged jet as previously
discussed.  This would produce ${\cal O}(50)$ events at SLHC, 
including detector efficiencies but no kinematic cuts, which should
be highly efficient given that the $b$ quarks would come from the
decays of very heavy states.

\bigskip
\underline{\it AMSB scenario}
\smallskip

Another interesting SUSY scenario is that of anomaly-mediation (AMSB),
which typically possesses the characteristic of very small mass
splitting between the LSP and NLSP, which are usually the lightest
neutralino and chargino, respectively.  The NLSP can exhibit dramatic
vertex displacements in the detector of many centimeters before
decaying.  Because of this, SM backgrounds would be non-existent and
the efficiency for capturing AMSB events would be extremely high,
close to $100\%$, as in some GMSB scenarios.  Charginos would decay to
$e$ and $\mu$ each about 1/7 of the time, giving an overall clean,
charge-identifiable state about $8\%$ of the time in stop-sbottom
events.

We do not explore AMSB parameter space widely, but find in general
that it is difficult to obtain light-enough squark masses for LHC
stop-sbottom production to be large enough to be useful, while at the
same time avoiding the LEP constraints on the chargino mass.


\section{Outlook}
\label{sec:concl}

Our conclusions from this initial study relevant for LHC/SLHC are
threefold.  First, it does appear that there is a path to separately
measuring weak production of mixed-flavor third-generation squark
pairs in many SUSY scenarios at the LHC/SLHC.  If further, more
detailed studies bear this out, it represents a significant increase
in the physics capability of LHC.  Second, the interpretation of weak
cross section measurements would depend strongly on what additional
information is extractable from other production processes.  The
signals proposed here could represent a way to help disentangle
third-generation squark mixing, at worst, or a potential measurement
of the weak vertex and the flavor matrix that diagonalizes squarks,
relative to the CKM matrix of the Standard Model.  This would depend
mostly what variety of SUSY is realized in nature, if at all: the
relative spectrum, the overall scale of SUSY masses, and other
unforeseeables.  While hints of generational squark mixing may show up
in QCD production channels, measuring the weak production vertex would
greatly aid in testing the flavor-diagonal hypothesis and measuring
deviations from it.  Third, any scenario which has a long-lived NLSP
is a boon for this measurement, as it would improve upon the detection
efficiency of the signal by (at a minimum) a factor 4-5.  Thus, any
given model with squark and gluino masses that would appear to have a
viable measurement of stop-sbottom production at SLHC would probably
become viable at LHC if it also had a long-lived NSLP.

Finally, we observe that our ability to pursue this interesting
physics is limited only by cross section.  Because squarks are
typically heavy, copious production at future colliders could
potentially open a new window onto these deeper questions, if SUSY is
found in nature.  One path is a next-generation (or beyond) linear
collider~\cite{ILC,Group:2004sz}, which could study the decays of
squark pairs with great precision to help disentangle the mixing
angles~\cite{Bartl:1997yi}.  Another possible path would be a
higher-energy hadron collider such as the proposed
VLHC~\cite{Baur:2002ka}, which would have the ability to produce
mixed-flavor squark pairs with cross sections typically two order of
magnitude larger than at LHC.


\subsection*{Acknowledgements}
We thank Tilman Plehn and Matt Strassler for highly useful discussions
and critiques, Howie Baer for suggestions about additional SUSY
backgrounds, and to Sally Dawson, Peter Zerwas and Lynne Orr for
reviews of the manuscript.  This research was supported in part by the
U.S. Department of Energy under grant No. DE-FG02-91ER40685.


\subsection*{Note added in proof:}
After submission of this work another paper appeared on the arXiv
which also calculates the $s$-channel stop-sbottom cross
sections~\cite{Bozzi:2005sy}.  Their calculations for this channel
agree with ours.


\end{document}